\documentclass[twocolumn,showpacs,preprintnumbers,amsmath,amssymb,prl]{revtex4}

\usepackage{graphicx}
\usepackage{dcolumn}
\usepackage{bm}
\usepackage{psfrag}

\usepackage{latexsym}

\usepackage{amsmath,amssymb,dsfont,mathrsfs}

\begin{document}

\preprint{}

\title{
Quantization of non-Abelian Berry phase for time reversal invariant systems
}

\author{Takahiro Fukui and Takanori Fujiwara}
 \affiliation{Department of Physics, Ibaraki University, Mito
310-8512, Japan}

\date{\today}

\begin{abstract}
We present a quantized non-Abelian Berry phase for time reversal invariant systems
such as quantum spin Hall effect. Ordinary Berry phase is defined 
by an integral of Berry's gauge potential along a loop (an integral of the Chern-Simons one-form), 
whereas we propose that a similar integral but over five dimensional parameter space
(an integral of the Chern-Simons five-form) is suitable to define a non-Abelian 
Berry phase. 
We study its global topological aspects and show that it is indeed quantized into two values. 
We also discuss its close relationship with the nonperturbative anomalies.

\end{abstract}


\maketitle

Quantum spin Hall (QSH) effect
\cite{MNZ03,Sin04,KMGA04,WKSJ04}
has attracted much renewed interest in topological order 
in condensed matter physics, providing us with a fundamental question about
its relationship with time reversal (${\cal T}$) symmetry.
For systems with broken ${\cal T}$ symmetry such as quantum Hall effect, 
the quantized Hall conductance is given by the TKNN integer \cite{TKNN82}, 
which is known to be the first Chern number \cite{TKNN82,Koh85} 
associated with the Berry phase \cite{Ber84} 
induced in the Brillouin zone \cite{Sim83}. 
It reflects the nontriviality of the U(1) Berry phase 
along loops in the two dimensional Brillouin zone. 
Namely, it is the sum of winding numbers around vortices which are
obstructions of the gauge-fixing of the wave functions.

On the other hand, for ${\cal T}$ invariant systems such as 
QSH systems, the first Chern number is always vanishing. 
This does not necessarily mean that ${\cal T}$ invariant insulators are
topologically trivial.
Recent findings are that the idea of the topological order is also 
crucial to the classification of ${\cal T}$ invariant insulating states. 
The Z$_2$ number proposed by Kane and Mele \cite{KanMel05a,KanMel05b}
for QSH effect \cite{KanMel05a,KanMel05b,BerZha06}
reveals that QSH phase is indeed a kind of topological insulator with ${\cal T}$ invariance.
More generically, the Z$_2$ number is a topological number specifying
${\cal T}$ invariant systems with ${\cal T}^2=-1$, where
${\cal T}$ denotes the time reversal transformation as well.

Some aspects of the Z$_2$ number in two dimensions (2D) have become clear and
its extension to 3D has been achieved 
\cite{XuMoo06,Roy06,SWSH06,FuKan06,FuKan07,FKM07,MooBal07,Roy06b,FukHat07,LeeRyu08,Mur07,TFK08}. 
Recent success in experimental observation 
of evidence of the topological insulating states 
in HgTe \cite{KWBetal07} and 
in Bi$_{1-x}$Sb$_x$ \cite{HsiQia08,HsiXia09,NisTas09}
urges us to clarify the topological meaning of the Z$_2$ number.
Towards this direction, it seems that topological properties in higher dimensional parameter space
play a crucial role \cite{ASSS88,QHZ08,FFH08}.
 
Another example of topological number for ${\cal T}$ invariant systems
is the quantized Berry phase proposed by Hatsugai \cite{Hat06}. 
While symmetry-broken phases are characterized by local order parameters,
those without symmetry-breaking, which occur especially in low dimensions, 
can be classified by topological order parameters.
The quantized Berry phase 
can serve as a local topological order parameter of gapped quantum liquids.
The quantized Berry phase in \cite{Ber84}, and the use of it as a 
local topological order parameter \cite{Hat06} was proposed only for ${\cal T}$ 
invariant systems with ${\cal T}^2=1$. Since it vanishes identically for ${\cal T}^2=-1$ systems, 
the generalization of the idea to the latter case 
is an intriguing problem both in its own right and with a view to clarifying the 
topological meaning of the Z$_2$ number.

In this paper, we propose a quantized non-Abelian Berry phase for ${\cal T}$ invariant
systems with ${\cal T}^2=-1$.
This can be achieved by the use of the Chern-Simons five-form, which shares several 
important features with the Chern-Simons one-form for the conventional Berry phase.
We show that it is indeed quantized into two values, and discuss the close
relationship with the nonperturbative anomalies in field theories of Weyl fermions \cite{Witten82}.

We first discuss a simpler case with ${\cal T}={\cal K}$, i.e., ${\cal T}^2=1$,
where ${\cal K}$ denotes the operator of taking complex conjugate. 
Let $H(x)$ be a time reversal invariant Hamiltonian, ${\cal T}H(x){\cal T}^{-1}=H(x)$,
where $x$ denotes a set of parameters $\{x_i\}$. 
In this case, Berry's gauge potential for the $n$ multiplet $\psi(x)$
is defined as $A=\psi^\dagger{\rm d}_x\psi$,
where ${\rm d}_x={\rm d}x_i\partial_{x_i}$ is the exterior derivative with respect to $x_i$.
Time reversal invariance ensures that ${\cal T}\psi(x)$ describes the same eigenstates of 
$H(x)$,  and hence, 
Berry's gauge potential for these ${\cal T}$-transformed wave functions is given by $A^*$.
It has been shown \cite{Hat06} that Berry phase, which is 
the integral of the gauge potential over a circle ${\rm S}^1$ 
in the space spanned by $x$, 
\begin{alignat}1
\Omega_1\equiv \frac{i}{2\pi}\int_{{\rm S}^1}\omega_1(A),
\label{QuaPha1}
\end{alignat} 
where $\omega_1(A)\equiv{\rm tr}\,A $ is the Chern-Simons one-form,
is quantized such that $\Omega_1=0$ or $1/2$ (in units of $2\pi$).
This can be proved by means of the following two properties:\\
(i) $\omega_1$ is gauge-dependent and its integral $\Omega_1$ can be defined 
only mod 1. In particular, since ${\cal T}\psi$ and $\psi$ denote the same eigenstates,
the difference between them is just a gauge transformation,
provided that $\psi$ is normalized.\\
(ii) $\omega_1$ is imaginary in the following sense: 
$\omega_1(A^*)=-\omega_1(A)$,  
which can be easily seen by noting that $A$ is anti-Hermitian and hence,
$A^*=-A^{T}$. \\
These two properties lead to the relationship $\Omega_1=-\Omega_1$ mod 1,
implying that $\Omega_1$ is quantized such that $\Omega_1=0$, $1/2$ mod 1 \cite{Hat06}.

Now, we would like to show that
similar techniques apply to the case ${\cal T}^2=-1$, and
therefore, the Z$_2$ invariant could be understood as a consequence of 
a quantized Berry phase. 
In this case, Berry's gauge potential is sp($n$)-valued, as will be discussed below, 
and therefore, Chern-Simons one-form is identically zero. 
This is true for non-Abelian gauge potential belonging to
semi-simple Lie algebras. 
In other words, the fundamental group of a gauge group G mentioned above is 
trivial, $\pi_1$(G)=0.
Therefore, the conventional Berry phase (\ref{QuaPha1})
plays no role in the case ${\cal T}^2=-1$.
To circumvent the difficulty, one can utilize 
higher forms for the present case.
Here, the property of the Chern-Simons ($2n-1$)-form ($n=1,2,\cdots$) 
\begin{alignat}1
\omega_{2n-1}(A^*)=(-)^{n}\omega_{2n-1}(A)
\label{CheSimReIm}
\end{alignat}
plays an important role. This tells that the Chern-Simons five-form,
\begin{alignat}1
\omega_5(A)={\rm tr}\,
\left[A({\rm d}A)^2+\frac{3}{2}A^3{\rm d}A+\frac{3}{5}A^5\right] ,
\nonumber
\end{alignat}
is also imaginary and hence, a natural candidate giving 
a quantized non-Abelian Berry phase.
As to the gauge dependence property (i) of the five-form, we have \cite{Zumino83}
\begin{alignat}1
\omega_{5}(A_g)=\omega_{5}(A)+{\rm d}\alpha_{4}(V_g,A)
+\frac{1}{10}{\rm  tr}\,(g^{-1}{\rm d}g)^5 ,
\label{GauDep}
\end{alignat}
where $A_g$ denotes the gauge transform of $A$, 
$A_g=g^{-1}Ag+g^{-1}{\rm d}g$, $V_g$ is a one-form defined by 
$V_g\equiv {\rm d}gg^{-1}$, and $\alpha_4$ is a four-form defined by  
\begin{alignat}1
\alpha_{4}(V,A)=&
-\frac{1}{2}{\rm tr}[V(AF+FA)]
\nonumber\\
&+\frac{1}{2}{\rm tr}\,[VA^3+V^3A+\frac{1}{2}(VA)^2] .
\end{alignat}
Here, $F$ is the field strength 2-form defined by $F={\rm d}A+A^2$.

These generic formulas are helpful to define a Berry phase
and to show it to be quantized.
Let $H(x)$ be a time reversal invariant Hamiltonian
${\cal T}H(x){\cal T}^{-1}=H(x)$, where ${\cal T}=i\sigma^2{\cal K}$ and 
we assume that $x$ stands for a set of five parameters
$x_i$ ($i=1,\cdots,5$), since the codimension of eigenvalue degeneracies 
is five \cite{ASSS88}. Let $\Psi(x)$ be a Kramers multiplet
with $2n$ degenerate states
\begin{alignat}1
\Psi(x)=\left(\psi(x),{\cal T}\psi(x)\right) .
\nonumber
\end{alignat}
Berry's gauge potential is defined by $A=\Psi^\dagger {\rm d}_x\Psi$, as usual.
Note that the
${\cal T}$ transformation yields ${\cal T}\Psi(x)=\Psi(x)J$ with $J=1_n\otimes i\tau^2$,
where $\tau^a$ is the Pauli matrix operating on the space of the Kramers doublet.
Then, one finds that the gauge potential $A$ belongs to sp($n$) algebra
because of the relation
\begin{alignat}1
A^*=JAJ^{-1} .
\nonumber
\end{alignat}
This equation also tells that the representation of $A$ is pseudo-real. 

Since the codimension of the eigenvalue degeneracies is five, we can choose in general
four dimensional sphere ${\rm S}^4$ on which there are no such dengeneracies 
except for the Kramers degeneracy. Inside ${\rm S}^4$, there could be a level-crossing
of several Kramers doublets, which yield an Sp($n$) monopole.
Consequently, the gauge potential may not be well-defined globally on S$^4$;
it can only be defined on several patches: For simplicity, we assume that 
two hemispheres ${\rm D}_\pm^4$ with ${\rm D}_+^4\cup{\rm D}_-^4={\rm S}^4$ 
are needed, and on ${\rm D}_\pm^4$ the gauge potential is given by $A_\pm$, respectively.
On the overlap region D$^4_+\cap$D$^4_-$, they are related each other as
\begin{alignat}1
A_+=h^{-1}A_- h+h^{-1}{\rm d}_xh ,
\nonumber
\end{alignat}
where $h(x)$ denotes a transition function satisfying
\begin{alignat}1
&
h(x)=Jh^*(x)J^{-1} ,
\nonumber
\end{alignat}
which, as well as the fact that $h$ is unitary by definition,
tells that $h$ is an element of Sp($n$).
The universality class is specified by the winding number of $h$ over ${\rm S}^3$,
since $\pi_3$(Sp($n$))=Z. This winding number is nothing but the second Chern number
discussed in \cite{ASSS88}.
Here, the difference of the Chern-Simons three-form 
$\omega_3(A_+)-\omega_3(A_-)$
is relevant to the integer winding number, 
but each three-form never gives a quantized Berry phase.

A natural integral domain for the Berry phase may be a compact manifold 
with dimension equal to its codimension minus one. 
In the previous case with ${\cal T}^2=1$, the codimension of eigenvalue degeneracies is two 
and the Berry phase is defined over ${\rm S}^1$ as described in Eq. (\ref{QuaPha1}), whereas
in the present case, the codimension is five and integral domain may be typically ${\rm S}^4$.
Therefore, the use of the five-form for a quantized Berry phase requires 
another degree of freedom.
This feature is analogous with the non-Abelian anomalies: 
These anomalies appear in four-dimensional gauge theories of Weyl fermions, but their
topological meaning 
can be revealed in five dimensions, 
where one extra dimension is a parameter of the gauge transformation 
which defines a gauge-orbit space \cite{AlvGin84}.

According to the techniques in the non-Abelian anomalies, 
we next introduce a one-parameter family of
gauge transformations $g(x,\theta)$, where
the parameter $\theta$ serves as an extra dimension.
Then, provided $g(x,2\pi)=g(x,0)$,
we can regard the gauge transformation $g$ as being defined on a five dimensional 
space ${\rm S}^4\times {\rm S}^1$.
The gauge-transformed potential is denoted as 
${\mathcal A}=g^{-1}Ag+g^{-1}{\rm d}g$,
where we have extended the exterior derivative ${\rm d}_x$ to 
${\rm d}={\rm d}_x+{\rm d}_\theta$.
The transition function between ${\cal A}_\pm$, which corresponds to the 
gauge transform of $A_\pm$, respectively, is given by  
$\tilde h\equiv g^{-1}hg$. 
Now let us define a Berry phase which is expected to serve as a topological 
order for the ${\cal T}$ invariant systems with ${\cal T}^2=-1$;
\begin{alignat}1
\Omega_5&=\frac{i^3}{3!(2\pi)^3}
\left[
\int_{{\rm S}^4\times{\rm S}^1} \hspace*{-3mm}\omega_5({\cal A})
-\int_{{\rm S}^3\times{\rm S}^1} \hspace{-3mm}\alpha_4(V_{\tilde h},{\cal A}_-)
\right].
\label{QuaPha5}
\end{alignat}
The expression of the first term is quite formal:
When one calculates this integral, one must divide ${\rm S}^4$ into the 
patches ${\rm D}_\pm^4$ and use well-defined gauge potential ${\cal A}_\pm$ there, respectively.
The additional second term is integrated over
the boundary $\partial{\rm D}_+^4\times{\rm S}^1={\rm S}^3\times{\rm S}^1$.
Therefore, this term may be interpreted as a boundary term 
which is necessary for the phase to be quantized. 
We will show in several steps that the phase $\Omega_5$ is quantized 
for ${\cal T}$ invariant systems in the similar manner as $\Omega_1$, and therefore, 
can be regarded as a non-Abelian version of Eq. (\ref{QuaPha1}).

Firstly, it should be noted that $\Omega_5$ vanishes if 
$g$ is kept strictly Sp($n$)-valued \cite{footnote}.
This is not surprising, because 
the Chern-Simons five-form has intimate relationship
with the gauge anomalies, and it is well-known that 
the latter vanish for pseudo-real gauge potentials.
However, this never means that the present gauge potential is topologically trivial:
The SU(2) anomaly  \cite{Wit83a,Wit83b}, 
or more generically, non-perturbative anomalies \cite{EliNai84} cannot be 
described by local fields like Eq. (\ref{QuaPha5}), but
by the use of the following techniques of embedding,
their topologically distinct sectors
can be revealed only by local gauge fields.
To be concrete, the gauge transformation $g \in$ Sp($n$) is
embedded in SU($2n+1$), and the gauge potential ${\cal A}$ is likewise
\cite{Wit83a,Wit83b,EliNai84}. 
Note here that the symmetry of the transition function 
$h$ plays an important role in the classification 
of the universality class, so we impose the boundary condition that $\tilde h\in$ Sp($n$) 
on ${\rm S}^3\times{\rm S}^1$ when embedding $g$ into SU($2n+1$).
This does not necessarily mean that $g\in$ Sp($n$) on the boundary:
We assume that $g$ is factorized on the boundary ${\rm S}^3\times{\rm S}^1$ such that 
\begin{alignat}1
g(x,\theta)=h_0(x,\theta)g_0(x,\theta) ,
\label{FacG}
\end{alignat} 
where $h_0\in$ Sp($n$) and $g_0\in$ SU($2n+1$). Then, 
if $g_0$ is commutative with generic Sp($n$) elements, we have indeed $\tilde h\in$ Sp($n$).
Such $g_0$ is given by
$g_0=e^{i\theta\lambda/2}$ \cite{footnote2}, where $\lambda\in$ su($2n+1$) is 
\begin{alignat}1
\lambda={\rm diag}(\underbrace{1,1,\cdots,1}_{2n},-2n) .
\nonumber
\end{alignat}
The space denoted as $2n$ in the above is the space of Sp($n$).
The normalization of $\lambda$ should be  chosen such that 
$\tilde h(x,2\pi)=\tilde h(x,0)$ \cite{footnote2}.
Although the symmetry group Sp($n$) of the transition function is thus kept unchanged, 
the gauge potential itself is nontrivial even on ${\rm S}^3\times{\rm S}^1$, since
it has a nonzero $\theta$ component ${\cal A}_\theta$ 
through $g^{-1}{\rm d}_\theta g\in$ su($2n+1$).

Secondly, by making use of Eq. (\ref{GauDep}),  it turns out that 
Eq. (\ref{QuaPha5}) can be written as 
\begin{alignat}1
\Omega_5=\frac{i^3}{3!(2\pi)^3}
\int_{{\rm S}^3\times{\rm S}^1} \hspace*{-3mm}\widetilde\alpha_4(h,g;A)+n_g^{(5)} ,
\label{RewQuaPha5}
\end{alignat}
where $n_g^{(5)}$ is a winding number of $g$ over 
${\rm S}^4\times{\rm S}^1$ \cite{footnote3}, 
and the four-form $\tilde\alpha_4$ is defined by
\begin{alignat}1
\widetilde \alpha_4(h,g;A)&\equiv
\alpha_4(V_g,A_+)-\alpha_4(V_g,A_-)-\alpha_4(V_{\tilde h},{\cal A}_-)
\nonumber\\
&=\alpha_4(V_g,v_h)-\alpha_4(V_{\tilde h},v_g)-\alpha_4(V_h,A_-)+{\rm t.d.} 
\label{TilAlp4}
\end{alignat}
Here, the first line is the definition of $\widetilde\alpha_4$.
In the second line, 
$v_g\equiv g^{-1}{\rm d}g$, and t.d. $\equiv{\rm d}\beta_3$ with some three-form $\beta_3$
denotes the total divergence (exact form) which gives no contribution to Eq. (\ref{RewQuaPha5}).
The second line can be derived from a more basic decomposition formula,
\begin{alignat}1
\alpha_4(V_{gh},A)=\alpha_4(V_h,A_g)+\alpha_4(V_g,A)-\alpha_4(V_h,v_g)+{\rm t.d.}
\nonumber 
\end{alignat} 
Eq. (\ref{TilAlp4}) tells that $\tilde\alpha_4$ depends on the gauge potential $A$ 
only through $\alpha_4(V_h,A_-)$ as well as t.d. terms, both of which vanish by the integration
in Eq. (\ref{RewQuaPha5}) \cite{footnote4}.
Therefore, 
we dare to suppress the irrelevant $A$-dependence of $\widetilde \alpha_4$,
referring to it as $\widetilde\alpha_4(h,g)$ for simplicity.
It should be stressed that $\widetilde\alpha_4(h,g)$ can be nonzero if and only if 
$g$ depends on $\lambda\in$ su($2n+1$).

The expression (\ref{RewQuaPha5}) is helpful to show that $\Omega_5$ is indeed topological.
To see this, we note that under the variation of $h$ or $g$, we have
\begin{alignat}1
\delta\widetilde\alpha_4(h,g)=\frac{1}{2}{\rm tr}\,
\left(\Delta_{\tilde h}V_{\tilde h}^4-\Delta_{h}V_{h}^4 \right) +{\rm t.d.},
\label{DelAlp4}
\end{alignat}
where $\Delta_h\equiv \delta hh^{-1}$. 
It should be noted that there are no terms such as $V_g$;
$g$ appears only through $\tilde h\in$ Sp($n$).
Therefore, every form $V$ as well as the variation $\Delta$ 
in the above equation is strictly sp($n$)-valued, 
and hence, it vanishes identically.
The other term denoted by t.d. also vanishes due to the integration, of course.  
We thus conclude that $\delta\Omega_5=0$. 
This implies that $\Omega_5$ cannot change continuously.
In Eq. (\ref{FacG}), we have made a specific embedding of Sp($n$) into SU($2n+1$), 
but the observation here tells that as far as one imposes the boundary condition 
on $g$ such that $\tilde h\in$ Sp($n$) on ${\rm S}^3\times{\rm S}^1$, 
one can expect that $\Omega_5$ is quantized  for any other embeddings.

Finally, we show that $\Omega_5$ is indeed quantized under the present embedding
such that $\Omega_5=0$ or 1/2 mod 1.
To this end, note the following relations
\begin{alignat}1
&\widetilde\alpha_4(h,g_1g_2)=\widetilde\alpha_4(h,g_1)+\widetilde\alpha_4(g_1^{-1}hg_1,g_2)
+{\rm t.d},
\nonumber\\
&\widetilde\alpha_4(h_1h_2,g)=\widetilde\alpha_4(h_1,g)+\widetilde\alpha_4(h_2,g)+
\widetilde\beta_4(\tilde h_1,\tilde h_2) ,
\label{TilAlp4Dec}
\end{alignat}
where the last term $\widetilde\beta_4$ depends only on $\tilde h_j\in$ Sp($n$), and therefore,
it vanishes.
Applying Eq. (\ref{TilAlp4Dec}) to Eq. (\ref{TilAlp4}) with the assumption (\ref{FacG}), 
we have
\begin{alignat}1
\widetilde\alpha_4(h,g_0h_0)
&=\widetilde\alpha_4(h,g_0)+\widetilde\alpha_4(g_0^{-1}hg_0,h_0)
\nonumber\\
&=\alpha_4(i\lambda{\rm d}\theta/2,v_h)-\alpha_4(V_h,i\lambda{\rm d}\theta/2)
\nonumber\\
&=\frac{i{\rm d}\theta}{2}{\rm tr}\,v_h^3 .
\nonumber
\end{alignat}
Here, we have used the fact that 
the second term in the r.h.s. of the first line vanishes, 
since both $h_0$ and $g_0^{-1}hg_0\in$ Sp($n$). 
It thus turns out that $\Omega_5$ is given by
\begin{alignat}1
\Omega_5=\frac{1}{2}n_h^{(3)}+n_g^{(5)} ,
\nonumber
\end{alignat}
where $n_h^{(3)}$ is the winding number of the transition function $h$ over ${\rm S}^3$.
This equation tells that $\Omega_5$ is quantized $\Omega_5=0, 1/2$ mod 1, where mod 1
comes from $n_g^{(5)}$, i.e., the winding number of 
$g$ which is introduced artificially as a probe of specifying the topological sector
of Berry's gauge potential defined on ${\rm S}^4$.
It also claims the similarities between $\Omega_1$ and $\Omega_5$: 
The former (the latter) is basically
given by half the first (second) Chern number.

In summary, we have defined a quantized non-Abelian Berry phase for time reversal invariant systems
with ${\cal T}^2=-1$ by the use of the Chern-Simons five-form.  
Although we have concentrated on global aspects of the quantized non-Abelian Berry phase,
we would like to stress here that it can also serve as a local topological order parameter.
Namely, for ${\cal T}$ invariant models in any dimensions, we can compute a quantized 
non-Abelian Berry phase as a local order parameter by adding some suitable 
local perturbations \cite{Hat06}.

There still remains several important issues to be clarified.
In particular, the direct relationship between Z$_2$ number defined in 2D Brillouin zone
(${\rm T}^2$) \cite{KanMel05a,FuKan06} 
and the quantized Berry phase defined in a five 
dimensional parameter space (${\rm S}^4\times{\rm S}^1$) is not yet very clear. 
This is mainly due to the fact that the former is inevitably calculated through the
Fourier transformation which breaks time reversal symmetry in the sense that 
Kramers multiplet belong to opposite momentum sectors, 
whereas the existence of the Kramers degeneracy in the whole parameter space
plays a vital role in the present formulation of the quantized Berry phase.
We believe that 
if time reversal symmetry is broken, Sp($n$) monopole would split into $2n$
U(1) monopoles with opposite charges, which can be detected by the Z$_2$ number in the
Brillouin zone. However, the present formulation does not directly cover the 
theory of the Z$_2$ number. 

We would like to thank Y. Hatsugai and H. Oshima for fruitful discussions.
This work was supported in part by a Grant-in-Aid for Scientific Research
(No. 20340098).


\begin{thebibliography}{99}
\bibitem{footnote}
Eq. (\ref{CheSimReIm}) tells that Chern-Simons five-form is imaginary, 
whereas the Sp($n$) gauge potential is (pseudo-)real. 
Hence, Eq. (\ref{CheSimReIm}) directly means that $\Omega_5=0$ for 
Sp($n$) gauge potentials.
\bibitem{footnote2}
This $g_0(\theta)$ does not have the period $2\pi$.
In this case, by the use of $h_0$, $g=h_0g_0$ can recover the periodicity.
However, even without this periodicity of $g$, we can see from Eq. (\ref{DelAlp4})
that the variation includes only $\tilde h=g^{-1}hg$, 
and hence the $2\pi$-periodicity of $\tilde h$
is enough to regard the space spanned by $\theta$ as ${\rm S}^1$.
\bibitem{footnote3}
If we further impose the boundary condition $g(x,\theta=0)=1$,
we can regard $g$ as a mapping from ${\rm S}^5$ to SU($2n+1$),
which has a nontrivial winding number, since $\pi_5($SU($2n+1$))=Z.
\end{thebibliography}
\end{document}